# Process intensification for the production of the ethyl esters of volatile fatty acids using aluminium chloride hexahydrate as a catalyst

Luigi di Bitonto, Sandro Menegatti, Carlo Pastore*

Water Research Institute (IRSA), National Research Council (CNR), via F. de Blasio 5, 70132 Bari, Italy

*carlo.pastore@ba.irsa.cnr.it

**Abstract**

A new process for obtaining the ethyl esters of volatile fatty acids with ethanol by using aluminium chloride hexahydrate as a catalyst is proposed. Aluminium chloride not only exhibits good activity, composition equilibrium is achieved within 3–4 hours at 343 K, but also induces a phase separation with a convenient distribution of the components. In fact, more than 99 %wt of the ethyl esters, together with most of the unreacted acid and ethanol, were found in the upper layer, which was well separated from the bottom phase, which contained the co-formed water and over 97.8 %wt of the catalyst. The intensification of this reaction and separation was thoroughly investigated and the operational conditions optimised. The effects of this separation on the purification of the final ethyl esters is fully investigated. A new configuration of unit operations is designed for the specific production of ethyl acetate, simulated through Aspen Plus V9® and compared with the current industrial process based on sulfuric acid catalysis. The overall production and purification of ethyl acetate is economically competitive, reduces the energy requirements by more than 50 %, and is potentially a zero-waste process, resulting in cleaner production.

**1. Introduction**

Ethyl esters are non-hazardous organic compounds that have industrial applications as solvents (Hu et al., 2017), fragrances (Saerens et al., 2008), cosmetic products, (Lee et al., 2014) and biofuels (Koutinas et al., 2016). These naturally occurring compounds (fruit flavours) have low toxicity and very limited impact on the environment; they can be easily hydrolysed to ethanol and native acids, which are biodegradable either aerobically (Bernat et al., 2017) or anaerobically (Pagliano et al., 2017). In addition, ethyl esters are bio-derived solvents because they can be produced through the direct esterification of volatile fatty acids (VFAs) and ethanol, both of which can be obtained via the fermentation of renewable biomasses. The production of ethanol through fermentation is a mature technology (Sebayang et al., 2017) and has been optimised for several residual biomasses (Sebayang et al., 2016). Additionally, the production of VFAs is a highly flexible process, in which the desired profile can be achieved by selecting the appropriate operating conditions, such as the type of inoculum and the pH (Wang et al., 2014) or the total solid content (Forster-Carneiro et al., 2008). VFAs or ethanol may be produced from the same fermenter by simply adopting specific operating conditions (Syngiridis et al., 2014). The efficiency and viability of the recovery of VFAs from broad fermentation have been increasingly improved (Singhania et al., 2013). The use of bio-derived VFAs and ethanol in place of fossil sources could contribute to a slowdown of the net increase in greenhouse gases emissions due to the 'short-cycle carbon system' (Kajaste, 2014).

Once isolated, they can react through direct esterification to produce ethyl esters. This process, known as the Fischer reaction (Eq. 1), has been widely studied by academics and industry and is subject to severe kinetic and thermodynamic constraints.

$$RCOOH + C_2H_5OH \leftrightarrows RCOOC_2H_5 + H_2O \qquad (1)$$

$$R = CH_3; C_2H_5; C_3H_7$$

Regarding the kinetics, the presence of a catalyst (typically an acid) is preferable because auto-catalysed reactions through the autoprotolysis reaction of the organic acid are slow and thus not always adequate for industrial purposes (Aslam et al., 2010). Homogeneous mineral acids (sulfuric acid, hydrochloric acid) efficiently promote direct esterification and are therefore typically used in industrial contexts. However, despite their effectiveness and affordability, homogeneous mineral acids are hardly recoverable or re-usable (de la Iglesia et al., 2007).

Nevertheless, due to their highly reactive and corrosive conditions, reactors and pipelines need to be made with expensive, nonreactive materials (Lu et al., 2013). These properties not only negatively impact the economy of chemical plants, but also necessitate the implementation of strict health and safety procedures in the work environment. The separation of spent catalysts from the final mixture results in the co-production of waste (sodium or calcium sulphates) that needs to be disposed of at the end of the process. Therefore, alternative reactive systems are being studied and developed, with a preference towards heterogeneous catalysts (i.e. zeolites (Wu and Chen, 2004), earth oxide and alumina-promoted $SO_4^{2-}/ZrO_2$ (Yu et al., 2009), acid resins (Pappu et al., 2013), carbon nanotubes (Cho et al., 2018) and metal oxides (Liu et al., 2015)). These systems are preferred for their favourable separation, recoverability, and potential reusability at the end of a reactive cycle. For the same purpose, supported enzymes have also been investigated (Koutinas et al., 2018).

Regarding the thermodynamics, the Fischer reaction is a chemical equilibrium that is strongly dependent on i) the operating temperature, ii) the nature of the acid to be converted and iii) the reaction media (solvent) (Liu et al., 2006). To achieve high yields (>90%), extreme conditions of temperature and pressure are required for the esterification process (573 K, >1 atm), thus increasing the costs of its production and management (Lee et al. 2017).

There is only a partial conversion of acids to the relevant esters, and the recovery of pure products in an industrial context is complicated by the coexistence of unreacted acids, ethyl esters, water and ethanol in the crude homogeneous reaction mixture, which requires several further expensive unit operations for purification (Aslam et al., 2010). To simplify the recoverability of the products and to promote equilibrium versus higher conversion, the typically adopted approach consists of removing water from the reactive environment in agreement with the principles of process intensification (Stankiewicz and Moulijn, 2000). Each process that includes the integration of a reaction and a separation represents a typical case of a process-intensifying method. Reactive distillation (using self-crosslinking Nafion–$SiO_2$ (Deng et al., 2016), or acid ion-exchange resins (Smejkal et al., 2009)) and pervaporation (using a mordenite membrane (Zhu et al., 2016) or zeolites (Tanaka et al., 2001)) completely convert the starting acid to the corresponding ethyl ester in a relatively short time (4–10 h). In addition, microwave-assisted reactive distillation (Ding et al., 2016) and reactive distillation coupled with membrane pervaporation (Lv et al., 2012) also represent good alternatives with improved performance. The chemical sequestration of water, for example through dicyclohexylcarbodiimide (Sano et al., 2011), is also a valid alternative.

Most of these alternatives cannot compete with the present industrial process, especially because the final purification has not been evaluated.

Recently, aluminium chloride hexahydrate ($AlCl_3 \cdot 6H_2O$) was reported to be an active catalyst in the direct esterification of long chain free fatty acids and methanol to produce biodiesel (Pastore et al., 2014), even on waste cooking oil and animal fat (di Bitonto and Pastore, 2019). Furthermore, as AlCl$_3$·6H$_2$O remained mainly dissolved in the methanol phase, well separated by the biodiesel produced, and was completely recoverable and reusable in new reaction cycles (di Bitonto et al., 2016). AlCl$_3$·6H$_2$O is affordable, less aggressive than conventional mineral acids, and can be used after catalysis as a coagulant in primary sedimentation in wastewater treatment plants (WWTPs) (Lin et al., 2018) with the aim of recovering new resources (VFAs).

The use of AlCl$_3$·6H$_2$O results in process intensification because the promotion of the direct esterification of long chain free fatty acids and the effective separation of the co-produced water from the reaction occurrs simultaneously through dissolution into the alcoholic phase (Pastore et al., 2015).

In this work, AlCl$_3$·6H$_2$O was tested as a catalyst for the direct esterification of VFAs with ethanol. Specifically, the reactions of ethanol with acetic (AA), propionic (PA) and butyric (BA) acids have been investigated, and the resulting kinetic ($E_a$ and $k_1$) and thermodynamic ($\Delta H^0$, $\Delta S^0$ and $K_{eq}$) parameters determined. AlCl$_3$·6H$_2$O was not only active in promoting direct esterification on par with mineral acids, but also able to induce a concomitant separation of the ethyl esters of VFAs from the co-formed water. The effects of the conditions of the catalysis (temperature, VFAs to ethanol molar ratio and amount of catalyst) on the VFA conversion and phase repartition were assessed and optimised to maximise both conversion and repartition. Consequently, the benefits of using AlCl$_3$·6H$_2$O have been thoroughly evaluated, particularly regarding the purification of the final ethyl esters. The phase separation establishes the potential for a new industrial process as an alternative to the conventional sulfuric acid-based system, which can be studied with the aim of obtaining ethyl acetate (EA) as a pure product. To date, over three million (MM) tons of EA have been produced worldwide, most generated by using sulfuric acid as a catalyst through a conventional process (Santaella et al., 2015). AA was chosen because it represents the most bio-available among the VFAs in fermentation (Wang et al., 2014; Forster-Carneiro et al., 2008) and therefore showed potential for obtaining EA as biobased solvent (Singhania et al., 2013). Through the use of a simulation program (Aspen Plus V9$^®$), the dimensioning the principal equipment involved in the proposed purification scheme as well as the production costs, energy intensity, conversion, recovery, Sheldon factor and mass intensity have been calculated and compared with the corresponding data for conventional industrial production reported by Santaella et al. (2015).

## 2. Materials and Methods

All chemical reagents used in this work were of analytical reagent grade and were used directly without further purification or treatment. Aluminum chloride hexahydrate ($AlCl_3·6H_2O$, 99 %) was purchased from Baker. Acetic acid ($CH_3COOH$, 99.5 %), propionic acid ($C_2H_5COOH$, ≥ 99.5 %), butyric acid ($C_3H_7COOH$, ≥ 99 %), ethyl acetate ($CH_3COOC_2H_5$, ≥ 99.8 %), ethyl propionate ($C_2H_5COOC_2H_5$, ≥ 99.5 %), ethyl butyrate ($C_3H_7COOC_2H_5$, ≥ 99.5 %), ethanol ($C_2H_5OH$, ≥ 99.9 %), hydrochloric acid (HCl, 37 %), sulfuric acid ($H_2SO_4$, 98 %) and p-toluen-sulfonic acid monohydrate ($CH_3C_6H_4SO_3H·H_2O$, ≥ 98.5 %) were purchased from Carlo Erba.

Qualitative identifications of chemical species were carried out by using a Perkin Elmer Clarus 500 gas chromatograph interfaced with a Clarus 500 spectrometer (GC-MS). Gas chromatographic quantitative determinations of ethyl esters and residual ethanol were performed by using a Varian 3800 GC-FID and ethyl benzene ($C_6H_5C_2H_5$, ≥ 99.5 % Sigma-Aldrich) as internal standard, using calibration curve prepared with EA, EP and EB as pure standards. Both instruments were configured for split injection with a HP-5MS capillary column (30 m; Ø 0.32 mm; 0.25 µm film). In detail, 1 µL of sample was injected in split mode (split ratio 1:3); helium was used as a carrier gas, with a flow of 2.8 mL min$^{-1}$. The temperature of the injection port was set at 523 K. Initial oven temperature was set to 313 K, and it was kept constant for 2 min. Then, the temperature was increased to 553 K (rate of increase 10 K min$^{-1}$) and held to the final temperature for 20 min. The temperature of detector (FID) was set to 573 K. Conversion of VFAs (acetic, propionic and butyric acid) was determined by titration of the residual acidity of the samples collected with a 0.1 N KOH solution (Aldrich) and phenolphthalein (≥ 99 %, Sigma-Aldrich) as indicator (di Bitonto et al., 2016).

Aluminum analysis of the phases recovered at the end of the esterification process were carried out using a 7000X ICP-MS instrument (Agilent Technologies). 0.1 g of sample were suspended in 9 mL of HCl, 3 mL of $HNO_3$, 4 mL of $H_2O_2$ and heated for 2 h at 503 K using a microwave oven (Milestone START E). Then, the mineralized samples were suspended into 100 mL of Milli-Q water and analyzed (ASTM D857-17).

Chloride analysis were performed by titration with a 0.1 N $AgNO_3$ solution (Sigma-Aldrich) and potassium dichromate ($K_2Cr_2O_7$, ≥ 99 % Sigma-Aldrich) as indicator (ISO 9297, 1989).

*2.1. Direct esterification of VFAs with ethanol using $AlCl_3 \cdot 6H_2O$ as a catalyst*

The direct esterification reaction of VFAs with ethanol was carried out in a glass reactor equipped with a silicone cap, which allowed sampling throughout the reaction without interruption, agitation, or heating the system. AA, PA, or BA were introduced into the reactor with ethanol and placed into a thermostatic oil bath (343, 333, 323 and 313 K) and magnetically stirred (250 rpm). Then, a previously prepared ethanolic solution of $AlCl_3·6H_2O$ was introduced via syringe into the reactor, to obtain the final acid:ethanol:catalyst molar ratio required for the specific experiment. Samples (0.2 mL) were collected at 30, 60, 90, 120, 150, 180, 240 and 480 minutes and analysed for any residual acidity and ethyl ester. At the end of the esterification process, when a bi-phasic system was observed, the two distinguishable phases were recovered, weighed and analysed for residual acids, ethyl ester, ethanol, water, aluminium and chloride content. Experiments were performed in triplicate for exhaustive treatment of the data (evaluation of the mean value and the respective error, which always resulted to be within 5 %).

*2.2. Phase repartition in the esterification of AA with ethanol*

The effect of the amount of catalyst on the phase repartition was evaluated on a synthetic mixture with a known thermodynamic composition obtained by reacting an equimolar mixture of AA and ethanol (343 K, 8 hours). In a glass reactor, 3.52 g AA was combined with 2.7 g ethanol, 11.4 g ethyl acetate (EA) and 2.34 g water. The resulting solution was a homogenous system in which no phase separation was observed. Then, 0.45 g $AlCl_3·6H_2O$ (1 % mol of starting AA used in the esterification process) was added to form a bi-phasic system. The two phases were recovered, weighed and analysed for AA, EA, ethanol, water, aluminium and chloride content. Finally, a systematic study was conducted to evaluate the catalytic effect of loading varying amounts of $AlCl_3·6H_2O$ (2, 3, 4 and 5 %mol). The phase repartition in the study was compared with that for HCl, $H_2SO_4$, and p-toluene-sulfonic acid under the same experimental conditions.

*2.3. Purification of EA: Process modelling and the optimisation method*

Industrial production of EA is nowadays conducted in large plants that have a capacity for manufacturing around 100 000 t of products per year using $H_2SO_4$ as a catalyst. The conventional scheme of production of EA reported by Santaella et al. (2015) was considered as the reference case in this study. In order to directly compare this conventional production with the process based on the use of $AlCl_3·6H_2O$ as a catalyst, a final EA production capacity of 12 255 kg per hour (8 160 hours per year) was selected.

The composition of the feed was the input data: the chemical composition of the organic layer obtained at the end of the esterification process using 5 %mol catalyst was used. The purification process was designed by considering a first distillation of the reacted mixture with the aim of separating EA from the residual AA (DC1), followed by an extractive distillation of the distillate using dimethyl sulfoxide (DMSO) (Zhang et al., 2018) which consisted in two further columns (namely EC and DC2). The total number of plates, the feeding plate, the distillate flow and the reflux ratio were the independent variables (factors) for all the columns and were iteratively varied to obtain the best combination that satisfied the specific separation criteria defined for each column and had the minimum energy. More precisely, in the first distillation, the complete recovery of EA and the maximum purification of AA were the target objectives, while the purity and recovery completeness of EA and EtOH were considered in EC and DC2. The range of variability for the different factors for the design specifications of the distillation columns are listed in Table 1.

**Table 1**

The thermodynamic non-random two-liquid equation was used to predict the physico-chemical properties of the chemical components involved in the distillation processes (Kenig et al., 2001). All sequences were modelled and simulated using Aspen Plus V9® (using the RadFrac column module). To optimise the conditions for the recovery of EA from the reaction mixture, a stochastic optimisation method was used (differential evolution with tabu list; Srinivas and Rangaiah, 2007). The process was improved using a hybrid platform of Microsoft – Aspen Plus V9®. The vector of design variables was sent from Microsoft Excel to Aspen Plus using Dynamic Data Exchange through COM technology. When the simulation was complete, the output from Aspen Plus is a Microsoft Excel file with the resulting vector that analyses the results and proposes new values for the decision variables.

2.4. Definition of the sustainability indicators

After the optimisation procedure, the sustainability indicators were determined to conduct a comparison of the entire process. The conversion (C), recovery (Rc) and productivity (P) were calculated by using Eqs 2–4 with respect to the two reactants (Re: EtOH and AA):

$$C(Re) = \frac{\text{Moles of Re converted}}{\text{Moles of Re fed}} \quad (2)$$

$$\text{Rc (Re)} = \frac{\text{Moles of EA in product stream}}{\text{Moles of Re converted}} \tag{3}$$

$$\text{P (Re)} = C \cdot Rc = \frac{\text{Moles of EA in product stream}}{\text{Moles of Re fed}} \tag{4}$$

These indicators contribute to the evaluation of the inherent safety and therefore the sustainability of the proposed process because the conversion, recovery, productivity and yield are directly related to the inventory of the reactants and the recycling streams flow rates.

Next, the energy intensity (EI), Sheldon's factor (E), water-free Sheldon's factor ($E_w$) (Sheldon, 2000), mass intensity (MI) and mass productivity (MP) (Jimenez-Gonzalez and Constable, 2011), were determined according to Eqs 5–9.

$$\text{EI} = \frac{\text{Energy used (W)}}{\text{Mass of product (kg)}} \tag{5}$$

EI represents the amount of energy used per kilogram (kg) of pure product. In this study, we considered the major sources of energy consumption to derive from the distillation processes (Santaella et al., 2015).

$$E = \frac{\text{Total waste streams (kg)}}{\text{Mass of product (kg)}} \tag{6}$$

$$E_w = \frac{\text{Total mass stream (kg)} - \text{Water in waste stream (kg)}}{\text{Mass of product (kg)}} \tag{7}$$

The E factor is an immediate measure of the amount of waste generated per kg of product, while $E_w$ does not include the water in the waste evaluation.

$$\text{MI} = \frac{\text{Total mass fed as pure reactants (kg)}}{\text{Mass of product (kg)}} \tag{8}$$

$$\text{MP} = \frac{1}{\text{MI}} \cdot 100 \tag{9}$$

The MI factor represents the amount of reagent required to synthesise one kg of the desired product (taking into account the eventual presence of water and excluding it from the computation). This factor is equal to 1 in the cleanest processes, in which the reagents are completely converted to useful products. The greater the MI factor is, the greater the amount of waste produced.

Finally, the MP factor is the inverse of the MI, and represents the mass of the reagent (percentage) converted to products.

These indicators provide an immediate measure of the cleanness of a process in accordance with the principles of green chemistry in terms of waste generated and energy efficiency (Anastas and Eghbali, 2010).

After the simulations met the design criteria, the total annual costs (TAC) were computed considering a 3-year period for return on the investment. Fixed costs were calculated using the method proposed by Douglas (1988) (Eqs 10–13). To calculate the variable costs, the average raw material and utility prices recently reported have been consulted (Santaella et al., 2015). Natural gas was used as the fuel, and an 85 % efficiency was assumed for the heating loop.

$$\text{TAC} = \text{Fixed Costs} + \text{Variable Costs} \qquad (10)$$

$$\text{Fixed Costs} = \frac{\text{Installed Costs}}{3 \text{year}} \qquad (11)$$

$$\text{Installed Costs} = (\text{Base Cost})(\text{Cost index})(\text{IF} + \text{Fc} - 1) \qquad (12)$$

IF is the installation factor, and Fc is a correction factor for materials, pressure, etc. The operating costs were calculated based on the consumption of utilities, specifically the heating costs.

$$\text{Energy Costs}\left(\frac{\text{USD}}{\text{year}}\right)$$

$$= \frac{\text{Heat duty}\left(\frac{\text{kW}}{\text{h}}\right) \cdot \text{Natural gas price}\left(\frac{\text{USD}}{\text{m}^3}\right) \cdot 8160\left(\frac{\text{h}}{\text{year}}\right)}{0.85 \cdot \text{Natural gas energy}\left(\frac{\text{kW}}{\text{m}^3}\right)} \quad (13)$$

## 3. Results and Discussion

*3.1. Direct esterification of VFAs and ethanol mediated by aluminium chloride hexahydrate*

Pure AA, PA and BA were reacted with a stoichiometric amount of ethanol in a closed glass reactor at different temperatures (313, 323, 333 and 343 K) in the presence of catalytic amounts of $AlCl_3 \cdot 6H_2O$ (1 %mol with respect to the starting acids) (Fig. 1).

**Fig. 1**

The direct esterification was monitored in time (for 8 hours) by analysing the residual acidity and the corresponding ethyl esters (in all cases, both values were congruent). According to the literature (Zhu et al., 2016), sulfuric acid requires approximately 4 h to reach equilibrium; therefore the final reaction time was 8 h. The reactive trends are reported in Fig. 2.

**Fig. 2**

The experiments were repeated three times, and the respective error bars for each set of data were calculated and represented. The variability of the experimental data was very small (less than 5 %).

The kinetic profiles in Fig. 2 suggest the following points: i) there is a positive effect of temperature on the kinetics and thermodynamics of the reaction, and an increase in temperature improves the rate of the reaction and the final conversion to esters; ii) the kinetics and thermodynamics of direct esterification strongly depend on the nature of the reacting acid in that once they are fixed, the temperature and EtOH:acid molar ratio (*r*), reaction rate and final equilibrium composition follow the order AA > PA > BA in relation to the size of the alkyl tail of the carboxylic acid used, in agreement with previous studies (Liu et al., 2006); and iii) the effect of the presence of the catalyst is clear: in the absence of $AlCl_3·6H_2O$, the reaction occurred very slowly, since at 343 K after 8 h, the final molar conversions were 11.8, 4.8 and 1.7 % for AA, PA and BA, respectively. Based on these experimental data, a more specific kinetic elaboration was carried out by verifying the fitting of a second order model for a homogeneous reaction (Akyalçin and Altıokka, 2012):

$$v = \frac{d[RCOOH]}{dt} = k_1 \left( [RCOOH][C_2H_5OH] - \frac{[RCOOC_2H_5][H_2O]}{K_{eq}} \right) \quad (14)$$

where *v* is the reaction rate, and $k_1$ and $K_{eq}$ are the kinetic constants for the forward reaction and the equilibrium constant respectively, and the molar concentration for each component refers to the equilibrium state. For the cases in which all the experiments were conducted with *r* equal to 1, the differential equation (Eq. 14) can be solved by introducing the *Y* function (Akyalçin and Altıokka, 2012) defined as

$$Y = \frac{1}{2\left(\frac{1}{X_{eq}} - 1\right)[RCOOH]_{t_0}} \ln\left[\frac{X_{eq} - (2X_{eq} - 1)X_t}{X_{eq} - X_t}\right] = k_1 t \quad (15)$$

where $X_{eq}$, $X_t$ and $[RCOOH]_{t_0}$ represent the acid conversion at the time of equilibrium (more precisely, the experimental value of *X* evaluated at the reaction time of 8 hours was used), at time (*t*), and with the starting molar concentration of the organic acid, respectively. While $X_{eq}$, $X_t$ and $[RCOOH]_{t_0}$ were all experimentally determined variables (average values obtained from triplicates of the experiments were used), $k_1$ was graphically obtained by plotting $Y$ vs $t$, to give the slope for the linear fitting the data using the fitting equation in the form $Y = k\,t$ according to Eq. 15 (Fig. 3).

**Fig. 3**

The values of $k_1$ are listed in Table 2 together with the $K_{eq}$ calculated using Eq. 16.

$$K_{eq} = \frac{X_{eq}^2}{(1-X_{eq})^2} \qquad (16)$$

**Table 2**

The data in Table 2 suggest that $K_{eq}$ was strongly dependent on the acid; at equilibrium, the final AA conversion yields were higher than those obtained for PA and BA. For more specific information on the kinetics and thermodynamics of the direct esterification mediated by $AlCl_3\cdot 6H_2O$, the Arrhenius and Van't Hoff equations were applied (Eqs 17 and 18):

$$\ln(k_1) = \ln(A) - \frac{E_a}{R}\frac{1}{T} \qquad (17)$$

$$\ln(K_{eq}) = -\frac{\Delta H^0}{R}\frac{1}{T} + \frac{\Delta S^0}{R} \qquad (18)$$

where $T$ is the absolute temperature, $A$ is the pre-exponential factor, $E_a$ is the activation energy of the reaction, $R$ is the universal gas constant, $\Delta H°$ is the reaction enthalpy, or heat of the reaction, and $\Delta S°$ is the reaction entropy (Fig. 4).

**Fig. 4**

The results were collected and are listed in Table 3.

**Table 3**

The values of $E_a$ increased following the order AA > PA > BA (Table 3) because of increasing steric hindrance constraints. The absolute values of these reactions suggest lower values of $E_a$ than those calculated for heterogeneous catalysts (over 30 kJ K$^{-1}$ mol$^{-1}$) (Lu et al., 2014; JagadeeshBabu et al., 2011). This result confirms the higher efficiency of the homogenous catalysis. Regarding the thermodynamics, not only did the direct esterification result in an endothermic reaction that was favoured by heat and high temperature, but the $\Delta H^0$ and $\Delta S^0$ estimated in this context also matched previous determinations for the same reactions (JagadeeshBabu et al. 2011).

The effect of increasing the amount of AlCl$_3$·6H$_2$O in the direct esterification was also determined: when the catalyst concentration rose from 1 to 5 %mol, there was a clear improvement in the reaction kinetics (Fig. 5).

**Fig. 5**

At 343 K and in the presence of 5 %mol AlCl$_3$·6H$_2$O, the reactive system resulted in a bi-phasic equilibrium composition after only 15 minutes.

The effects of the amounts of AlCl$_3$·6H$_2$O on the rates of reaction were also extended to the PA and BA cases. The presence of more AlCl$_3$·6H$_2$O benefited the kinetics of the reactions of these two acids as well: in fact, the reactions occurred in less than 30 minutes.

To improve the conversion of the acids, the effect of $r$ was also investigated. In addition to the previously described studies in which $r$ was fixed at 1, the reactions in which $r$ was fixed at 2 and 3 were studied for AA, PA and BA under AlCl$_3$·6H$_2$O catalysis (Fig. 6).

**Fig. 6**

Although the final conversion of the initial acid increased with higher yields of the corresponding ethyl ester (conversions increased from 55%–66% for $r$ = 1, to 80%–82% for $r$ = 2, to 82%–85% for $r$ = 3), no phase separations were detected, even when increasing the amount of $AlCl_3·6H_2O$ to 5 %mol. With the increase in the value of $r$, the final conversions for the different acids were more similar than were those when the value of $r$ was 1. This effect could be due to the increasing presence of ethanol, which would influence the $K_{eq}$ of the reaction (Liu et al., 2006).

Finally, the direct use of azeotropic ethanol (ethanol:water = 96:4) as a reactant instead of absolute alcohol did not produce significant differences in the final conversion of the acids and the final separation of the phases.

*3.2. Effect of $AlCl_3·6H_2O$ on phase separation*

The catalysis of $AlCl_3·6H_2O$ with pure acids initially resulted in homogeneous solutions for AA, PA, and BA (Fig. 1c). In addition to the changes in the overall compositions due to the formation of the corresponding ethyl esters, bi-phasic systems were demonstrated in all the experiments in which the value of $r$ was fixed to 1 (Fig. 1d). For $r$ = 2 or 3, no separations occurred. To study and describe the bi-phasic system, the overall chemical composition was determined, and the quantification of the two different phases and the distribution of the different species among the two phases were monitored. For a given concentration of the acid and of the catalyst, the phase separation always occurred at the same overall composition, even when appearing at different temperatures (Table 4).

**Table 4**

The conversion necessary to generate the phase separation decreased with an increase in the length of the alkyl group of the acid (Table 4): in the presence of 1 %mol catalyst, the phase separation occurred at a conversion rate of 56.4 % for AA, whereas for PA and BA, the reactive mixtures became bi-phasic when 40.0 % and 30.2 %, respectively, of the starting acids were converted. After 8 hours of reaction, the resulting solutions were decanted to separate the two phases. These two phases were then weighed, and their constituent reaction products (ethyl ester and water) and residual reagents (acids and ethanol) were analysed. For the reactions carried out at 343 K in the presence of 1 %mol of $AlCl_3·6H_2O$, the denser phases were quantified as 7.2, 6.4 and 5.3 %wt for AA, PA and BA, respectively. In all these cases, effective separations were verified as the ethyl esters were completely dissolved in the upper phase, whereas the catalyst was mainly contained in the lower phase.

Next, an experiment to assess the effect of increasing the amount of $AlCl_3·6H_2O$ on the equilibrium phase composition was conducted for the case of AA. A mixture of AA (17.6 %wt), EA (57.2 %wt), water (11.7 %wt) and ethanol (13.5 %wt) was prepared, simulating the final equilibrium composition obtained from the reaction of an equimolar mixture of AA and ethanol at 343 K. This solution appeared to be homogeneous even after the addition of conventional mineral acids (HCl, $H_2SO_4$, p-toluene-sulfonic acid) at different catalyst to AA molar ratios (from 1 to 5 %). In contrast, when $AlCl_3·6H_2O$ was added at a low concentration (1 %mol), a separation of the phases was evident. Next, different catalyst amounts (ranging from 1 to 5 %mol) were added to the synthetic solution (Fig. 7), and the repartition of the phases and the final distribution of the different components were determined.

**Fig. 7**

Fig. 7 shows that increasing the amount of AlCl$_3$·6H$_2$O resulted in an increase in the lower phase from 7.2 %wt to 24.0 %wt. The AA, EtOH, EA, water and AlCl$_3$·6H$_2$O contents (Table 5) in the resulting phases were then analysed.

**Table 5**

According to the data reported in Table 5, EA was dissolved mainly in the upper phase (> 99 %wt), while AlCl$_3$·6H$_2$O was dissolved mainly in the lower aqueous phase. In addition, increasing the amount of the catalyst (up to 5 %mol), led to an increase in the amount of water in the lower phase, which resulted in an effective concomitant purification of EA and an almost complete dewatering of the upper organic phase, thus establishing a process intensification (reaction and separation of products in a single step).

*3.3. Advantages related to the use of AlCl$_3$·6H$_2$O instead of H$_2$SO$_4$ as the catalyst in producing EA*

The industrial production of EA is currently based on the application of the process shown in Fig. 8 (Santaella et al., 2015).

**Fig. 8**

The most challenging issue related to this industrial production is the downstream purification of the products, which plays a key role in the overall economy of the process. In this process, the reacted homogeneous mixture, composed of EtOH (3 473 kg h$^{-1}$), AA (16 086 kg h$^{-1}$), EA (35 131 kg h$^{-1}$) and water (5 889 kg h$^{-1}$), is first distilled (DC in Fig. 8) to obtain a distillate composed of the ternary azeotrope EtOH:EA:H$_2$O (having 0.1126:0.5789:0.3085 molar ratio) and a residue richer in AA (further purified through an azeotropic distillation (AD) and recycled back to the reactor). The simplest approach to break this ternary azeotrope and to recover pure EA is to produce two different phases by adding a large amount of water. Distillation (RC) of the organic layer generated by the addition of the water allows pure EA to be obtained as a residue. The remaining water, which contains large amounts of ethanol, EA and AA, needs further treatment and represents a waste product. The energy consumption correlated with the overall purification in this conventional process was calculated to be 26 582 kW. The critical step in this purification process is the separation of the ternary azeotrope, which can only be accomplished by adding a large amount of water.

Based on the results discussed in Section 3.2, $AlCl_3 \cdot 6H_2O$ not only promoted the direct esterification of EtOH and VFAs to produce the relevant ethyl esters, but also induced a concomitant effective separation of the co-obtained water into a different phase. This behaviour implies a drastic change in the downstream purification of EA. In fact, after reaction with 5 % $AlCl_3 \cdot 6H_2O$, the water was already separated and the purification of EA involved a simpler mixture, whose composition is reported in Table 5. For this reason, a new process can be designed and optimised through a simulation using Aspen Plus V9® (Fig. 9).

**Fig. 9**

In fact, considering that the purification involved the organic phase generated after the reaction, from which water was completely absent, (EtOH (2 542 kg $h^{-1}$), AA (3 543 kg $h^{-1}$), EA (12 268 kg $h^{-1}$), water (37 kg $h^{-1}$) and catalyst (37 kg $h^{-1}$)), the first distillation (DC1) produced a distillate composed mainly of EtOH and EA (17.1:82.5), as well as a residue of pure AA, which can be directly recycled back to the reactor. The distillate from DC1 appeared to be like an azeotropic EtOH:EA mixture, from which the purification of EA could be efficiently accomplished by extractive distillation (EC in Fig. 9) using DMSO (Zhang et al., 2018). Thus, pure EA can be distilled by EC, and the ethanol can be purely and quantitatively recovered through a third distillation (DC2). From the same distillation, DMSO can also be completely recovered and recycled back to the EC. While the energy required in the reaction was almost the same as that for the conventional process (330 kW), the overall energetic requirement (heating duty) for this purification process was calculated to be 9 780 kW, which is almost one-third that required for the conventional scheme of production.

To evaluate the practicability of the proposed process and to make possible a direct comparison with conventional production of EA, an economic feasibility test was carried out by considering the costs and method proposed by Santaella et al. (2015). Raw reagents, energy and fixed costs were estimated to be 98.76, 3.2 and 0.8 MM USD per year, respectively, confirming that the most important contribution to the determination of the value of the TAC (102.76 MM USD per year) is the raw material (> 95 %). The overall estimate of the TAC needs further adjustment due to the cost of the catalyst, but this factor was omitted and not considered in the conventional process.

Even a single run using aluminium chloride could be considered economically sustainable, because the total amount of catalyst needed corresponds to 20 000 t per year for an annual purchasing cost of 10.89 MM USD (Schwiderski and Kruse, 2016). Under these conditions, the final TAC is 113.65 MM USD per year, which is competitive with the conventional process, whose TAC is 132.3 MM USD per year.

To evaluate the benefits other than economic feasibility associated with the application of $AlCl_3 \cdot 6H_2O$ instead of sulfuric acid, a series of sustainability indicators were calculated and are reported in Table 6.

**Table 6**

All indicators demonstrate that the proposed process is cleaner than the conventional process. For conversion, recovery and productivity, both reactants were considered because they were used in stoichiometric amounts. It is clear that the most important difference occurred with EtOH due to its loss from the aqueous phase generated from the recovery of EA from the ternary azeotrope created with the addition of water in the conventional process (Fig. 8). The EI was also more advantageous because only one-third of the energy was required to sustain the proposed process. The estimated value of the MP factor was close to 0.83, which represents the theoretical maximum achievable for the direct esterification of ethanol and acetic acid (atom economy of the reaction).

Finally, less waste can be produced per kg of product (E), even when water is not included in the estimation of the generated waste ($E_w$).

Regarding the nature of the waste produced, the conventional process generates an aqueous stream that needs a very expensive treatment due to the presence of a very high concentration of organic compounds. The costs associated with this treatment are not included in the TAC, which results in an underestimation. In addition, sulfuric acid cannot be recycled many times, and new waste is generated, which needs to be disposed of. In contrast, the process based on the use of aluminium chloride generates only one highly contained waste stream (the E factor is 5 times smaller than that for the conventional process). Furthermore, this factor would be cancelled if the aqueous stream of aluminium chloride produced in the proposed process were to find a direct application in WWTP as a flocculant instead of the polychlorides of aluminium.

Under these conditions, the proposed scheme would not only be a potential zero-waste process (E = 0 and perfectly addressing the principles of green chemistry), but it could also be more economically advantageous because it could be sold to WWTPs.

*4. Conclusions*

In this work, $AlCl_3·6H_2O$ was proposed as a catalyst in the direct esterification of VFAs with ethanol to produce ethyl esters and to promote an effective separation of products from water. The effect of the nature of the carboxylic acid in the esterification process was investigated by collecting kinetic and thermodynamic data for acetic, propionic and butyric acids. The order of reactivity observed (AA > PA > BA) is related to the size of the carboxylic acids, with an evident reduction in the yields with the increase in the length of the alkyl group. The calculated $E_a$ was lower than the values determined for heterogeneous catalysts (> 30 kJ $K^{-1}$ $mol^{-1}$), confirming the higher efficiency of the process. In contrast with conventional mineral acids (HCl, $H_2SO_4$, p-toluen-sulfonic acid), $AlCl_3·6H_2O$ induces a favourable final separation of ethyl esters (> 99 %wt) from the co-formed water in two distinct phases. The starting load of the catalyst plays a key role in the kinetics and in the final separation of phases: with 5 %mol $AlCl_3·6H_2O$, the reaction reached equilibrium within 15–30 minutes, and there was an increase in the water content in the lower phase, which resulted in complete dewatering the organic phase.

To evaluate the main advantages associated with the use of $AlCl_3 \cdot 6H_2O$, a new process scheme for the production and purification of EA was proposed, simulated using Aspen Plus® and compared with the conventional process. A simplification of the purification process was achieved, and based on an annual production of 100 000 t of pure EA, the proposed system is not only economically advantageous, with a TAC of 113.65 instead of 132.3 MM USD per year, but it would also produce one-fifth of the waste by consuming one-third of the energy. In addition, taking into consideration that $AlCl_3 \cdot 6H_2O$ was effectively recoverable in an aqueous phase, which could potentially be used in WWTPs as a coagulant, cogeneration of waste could be eliminated, resulting in a zero-waste process.

All these factors cause the proposed technology to be competitive with the present conventional industrial process for the production of the ethyl esters of VFAs, thus fully satisfying sustainability criteria.

**Acknowledgements**

This work was supported by the REsources from URban BIo-waSte" - RES URBIS (Grant Agreement 730349) project in the European Horizon2020 (Call CIRC-05-2016) program.

**Abbreviations**

*Roman Letters*

$A$ = Pre-exponential factor (min$^{-1}$)

AA = Acetic acid

AD = Azeotropic distillation column

$AlCl_3 \cdot 6H_2O$ = Aluminum chloride hexahydrate

BA = Butyric acid

C = Conversion (%)

DC = Distillation column

DMSO = dimethylsulfoxide

E = Sheldon's factor

EA = Ethyl acetate

$E_a$ = Activation energy (kJ K$^{-1}$ mol$^{-1}$)

EB = Ethyl butyrate

EC = Extractive column

EI = Energy intensity (W kg$^{-1}$)

EP = Ethyl propionate

EtOH = Ethanol

$E_w$ = Water-free Sheldon's factor

Fc = Correction factor

h = Hour

HCl = Hydrochloric acid

$H_2SO_4$ = Sulfuric acid

IF = Installation factor

$K_{eq}$ = Equilibrium constant

kg = Kilogram

$k_1$ = Kinetic constant (L mol$^{-1}$ min$^{-1}$)

MI = Mass intensity

min = Minutes

MM = Million

MP = Mass productivity

P = Productivity

PA = Propionic acid

$R$ = Universal gas constant (J mol$^{-1}$ K$^{-1}$)

$r$ = Initial molar ethanol:acid ratio

Re = Reactant (AA or EtOH)

Rc = Recovery

[RCOOH], [RCOOC$_2$H$_5$], [C$_2$H$_5$OH], [H$_2$O] = Concentrations of acid, ethyl ester ethanol and water (mol L$^{-1}$)

T = Temperature (K)

$t$ = Time (min)

TAC = Total annual costs (MM USD year$^{-1}$)

VFAs = Volatile fatty acids

W = Watt

$X_t$, $X_{eq}$ = Conversions of acid at time (t) and equilibrium

*Greek Letters*

$\Delta H°$ = Reaction enthalpy (kJ mol$^{-1}$)

$\Delta S°$ = Reaction entropy (J K$^{-1}$ mol$^{-1}$)

$v$ = Reaction rate (mol L$^{-1}$ min$^{-1}$)

**Table 1.** Design specifications values and ranges for each column. DC1 = distillation column 1, DC2 = distillation column 2, EC = extractive column for EA recovery.

|  | Design specification (Intervals) | | |
|---|---|---|---|
| **Column** | **DC1** | **EC** | **DC2** |
| Pressure (atm) | 1 | 1 | 1 |
| Type of Stages | Bubble cap | Bubble cap | Bubble cap |
| Number of Stages | 15-25 | 15-30 | 15-25 |
| DMSO Stage |  | 2-12 |  |
| Feed Stage | 2-24 | 8-29 | 2-24 |
| Reflux ratio | 1.2-5 | 1.2-5 | 1.2-5 |

**Table 2.** $k_1$ and $K_{eq}$ determined for acetic, propionic and butyric acids at 313, 323, 333 and 343 K.

| T (K) | Acetic acid | | Propionic acid | | Butyric acid | |
|---|---|---|---|---|---|---|
| | $k_1 \times 10^{-3}$ (L mol$^{-1}$min$^{-1}$) | $K_{eq}$ | $k_1 \times 10^{-3}$ (L mol$^{-1}$min$^{-1}$) | $K_{eq}$ | $k_1 \times 10^{-3}$ (L mol$^{-1}$min$^{-1}$) | $K_{eq}$ |
| 313 | 0.83 ± 0.03 | 1.91 ± 0.05 | 0.74 ± 0.02 | 0.94 ± 0.02 | 0.41 ± 0.02 | 0.44 ± 0.01 |
| 323 | 1.20 ± 0.04 | 3.03 ± 0.03 | 1.09 ± 0.04 | 1.28 ± 0.02 | 0.66 ± 0.02 | 0.72 ± 0.01 |
| 333 | 1.39 ± 0.05 | 4.27 ± 0.08 | 1.34 ± 0.03 | 1.92 ± 0.06 | 1.00 ± 0.02 | 1.15 ± 0.02 |
| 343 | 1.81 ± 0.04 | 4.91 ± 0.11 | 1.60 ± 0.05 | 2.15 ± 0.04 | 1.36 ± 0.04 | 1.32 ± 0.03 |

**Table 3.** $E_a$, $\Delta H^0$ and $\Delta S^0$ calculated for the reaction of direct-esterification between acetic, propionic and butyric acid with ethanol, under $AlCl_3 \cdot 6H_2O$ catalysis.

| VFAs | $E_a$ kJ K$^{-1}$ mol$^{-1}$ | $\Delta H^0$ kJ mol$^{-1}$ | $\Delta S^0$ J K$^{-1}$ mol$^{-1}$ |
|---|---|---|---|
| Acetic Acid | 22.3 | 28.5 | 97.1 |
| Propionic Acid | 22.8 | 25.9 | 82.3 |
| Butyric Acid | 35.8 | 34.7 | 104.3 |

**Table 4.** Molar conversion of the starting acid at which separation of phases occurred for the three different acids and weight composition of the respective overall systems (r =1; $AlCl_3·6H_2O$ = 1%mol).

| VFAs | Acetic | Propionic | Butyric |
|---|---|---|---|
| % Conversion | 56.4 ± 0.2 | 40.0 ± 0.2 | 30.2 ± 0.7 |
| **Chemical composition of the overall system** | | | |
| Acid (%wt) | 24.5 ± 0.1 | 36.9 ± 0.1 | 45.8 ± 0.5 |
| Ethyl ester (%wt) | 46.6 ± 0.2 | 33.9 ± 0.1 | 26.2 ± 0.6 |
| Ethanol (%wt) | 19.3 ± 0.1 | 23.2 ± 0.1 | 24.0 ± 0.3 |
| Water (%wt) | 9.6 ± 0.1 | 6.0 ± 0.2 | 4.0 ± 0.1 |

**Table 5.** Chemical distribution (%wt) of acetic acid, ethanol, ethyl acetate, water and $AlCl_3 \cdot 6H_2O$ among the two phases (upper and lower).

| Catalyst loaded | 1%mol | 2%mol | 3%mol | 4%mol | 5%mol |
|---|---|---|---|---|---|
| | | | Upper phase | | |
| Acetic acid (%wt) | 98.8 ± 0.2 | 96.7 ± 0.3 | 95.6 ± 0.1 | 95.0 ± 0.3 | 92.3 ± 0.1 |
| Ethanol (%wt) | 95.4 ± 0.1 | 92.8 ± 0.2 | 88.8 ± 0.2 | 87.4 ± 0.2 | 86.7 ± 0.2 |
| Ethyl acetate (%wt) | 99.7 ± 0.1 | 99.5 ± 0.1 | 99.3 ± 0.2 | 99.1 ± 0.1 | 98.9 ± 0.2 |
| Water (%wt) | 62.0 ± 0.2 | 39.9 ± 0.1 | 25.9 ± 0.2 | 11.5 ± 0.1 | 1.4 ± 0.1 |
| $AlCl_3 \cdot 6H_2O$ (%wt) | 16.5 ± 0.1 | 5.6 ± 0.1 | 3.3 ± 0.1 | 2.2 ± 0.1 | 1.2 ± 0.1 |
| | | | Lower phase | | |
| Acetic acid (%wt) | 1.2 ± 0.1 | 3.3 ± 0.1 | 4.4 ± 0.2 | 5.0 ± 0.1 | 7.7 ± 0.1 |
| Ethanol (%wt) | 4.6 ± 0.1 | 7.2 ± 0.1 | 11.2 ± 0.1 | 12.6 ± 0.1 | 13.3 ± 0.2 |
| Ethyl acetate (%wt) | 0.3 ± 0.1 | 0.5 ± 0.1 | 0.7 ± 0.1 | 0.9 ± 0.1 | 1.1 ± 0.1 |
| Water (%wt) | 38.0 ± 0.1 | 60.1 ± 0.1 | 74.1 ± 0.3 | 88.5 ± 0.2 | 98.6 ± 0.2 |
| $AlCl_3 \cdot 6H_2O$ (%wt) | 83.5 ± 0.2 | 94.4 ± 0.2 | 96.7 ± 0.1 | 97.8 ± 0.3 | 98.8 ± 0.2 |

**Table 6.** Sustainability indicators calculated for conventional production process and optimized process using aluminum chloride hexahydrate as catalyst.

| Process | Conversion AcOH/EtOH | Recovery AcOH/EtOH | Productivity AcOH/EtOH | MI | E | Ew | MP | EI |
|---|---|---|---|---|---|---|---|---|
| Conventional | 0.98/0.80 | 0.86/0.85 | 0.84/0.68 | 1.58 | 2.23 | 0.34 | 0.63 | 2.17 |
| $AlCl_3 \cdot 6H_2O$ | 0.98/0.96 | 0.98/0.98 | 0.96/0.94 | 1.29 | 0.47 | 0.26 | 0.77 | 0.79 |

**Figure captions**

**Fig. 1.** a) Reaction apparatus with the thermostatic bath and glass reactor; b) detail of the silicon cap of the reactor; c) initial homogeneous reaction mixture; d) two phases obtained after carrying out direct esterification with $AlCl_3·6H_2O$.

**Fig. 2.** Kinetic profiles of the direct esterification of a) acetic, b) propionic and c) butyric acids with ethanol at different temperatures. Reaction conditions: molar ratio ethanol:acid:$AlCl_3·6H_2O$=1:1:0.01, temperatures from 313 to 343 K, time = 8 h.

**Fig. 3.** Evaluation of the kinetic constants for the forward reaction ($k_1$) for a) acetic, b) propionic and c) butyric acids.

**Fig. 4.** Arrhenius (a) and van't Hoff (b) plots for the ethyl acetate, ethyl propionate and ethyl butyrate syntheses through direct esterification of the respective acids.

**Fig. 5.** Kinetic profiles of direct-esterification of acetic acid with ethanol at different catalyst concentrations. Reaction conditions: molar ratio ethanol:acid = 1; $AlCl_3·6H_2O$ from 1 to 5 %mol, temperature = 343 K, time = 8 h.

**Fig. 6.** Kinetic profiles of direct-esterification of a) acetic, b) propionic and c) butyric acids with ethanol at different molar ratio ethanol:acid ($r$ = 1, 2 and 3). Reaction conditions: $AlCl_3·6H_2O$ = 3 %mol, temperature = 343 K, time = 8 h.

**Fig. 7.** Effect of different molar percentages of $AlCl_3·6H_2O$ in the separation of the phases.

**Fig. 8.** Conventional EA production using sulfuric acid (DC = distillation column, AD = azeotropic distillation column, RC = recovery column of EA).

**Fig. 9.** Optimised process using aluminium chloride hexahydrate (DC1 = distillation column 1, DC2 = distillation column 2, EC = extractive column for EA recovery). The energy optimization procedure referred only to the distillation processes.

**Fig. 1**

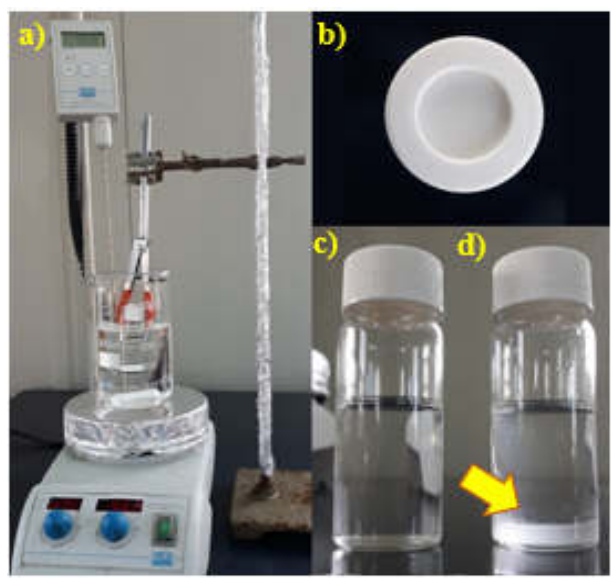

**Fig. 2**

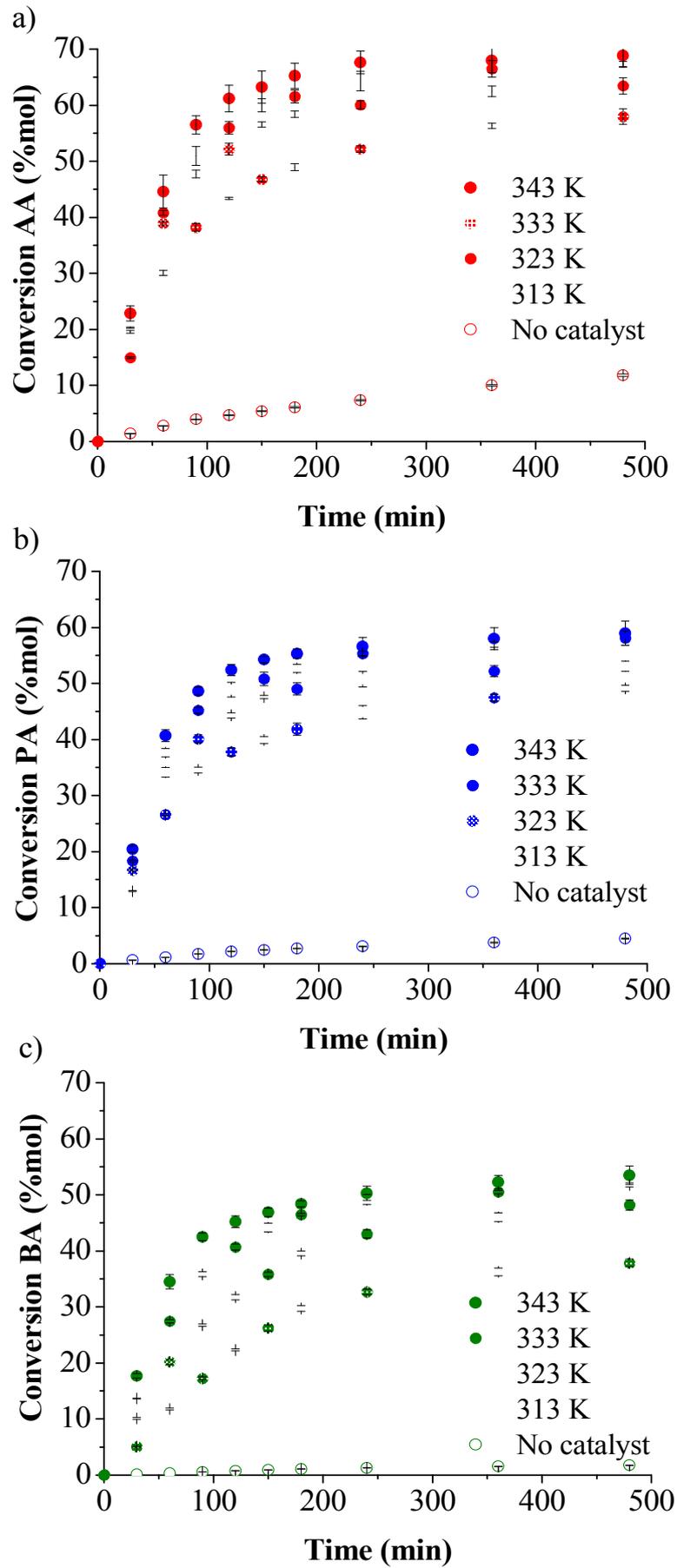

**Fig. 3**

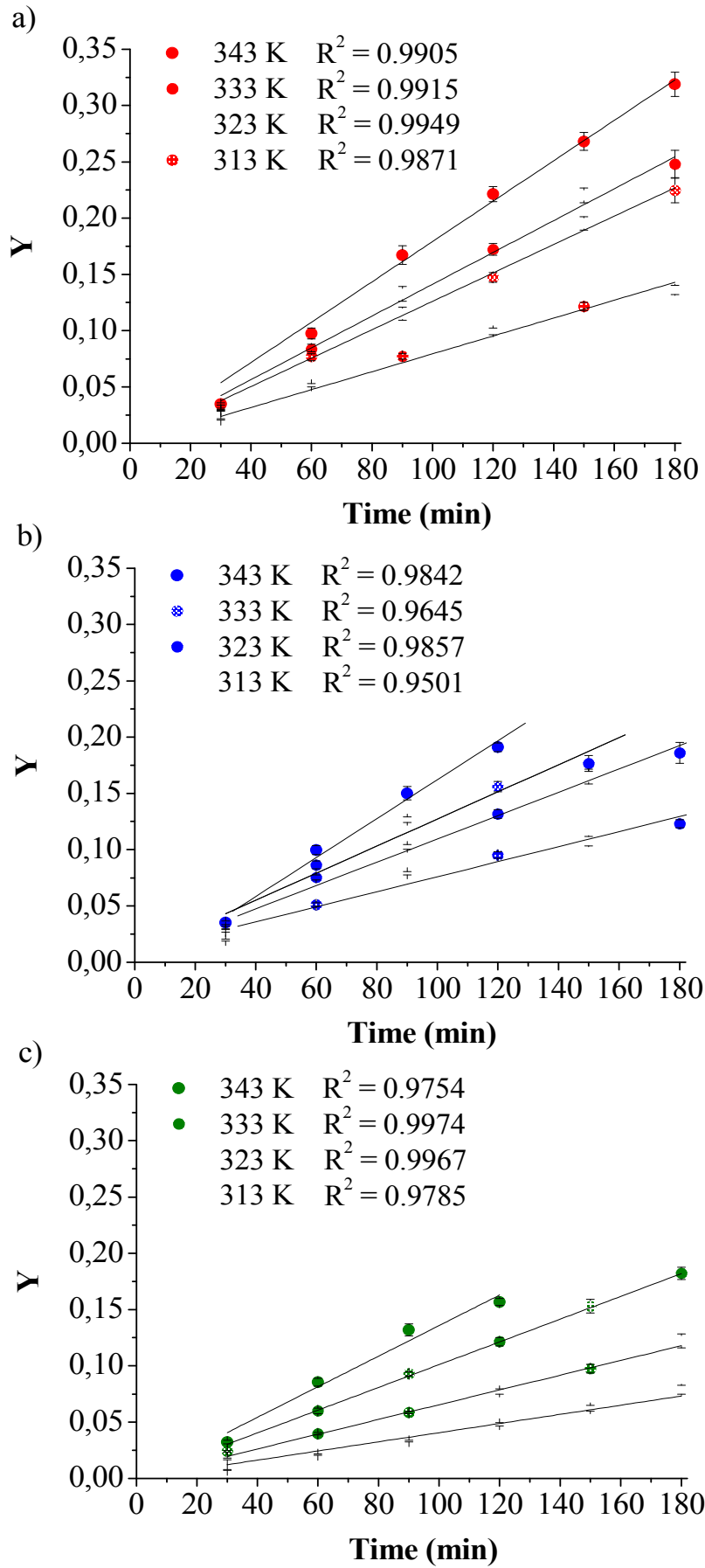

**Fig. 4**

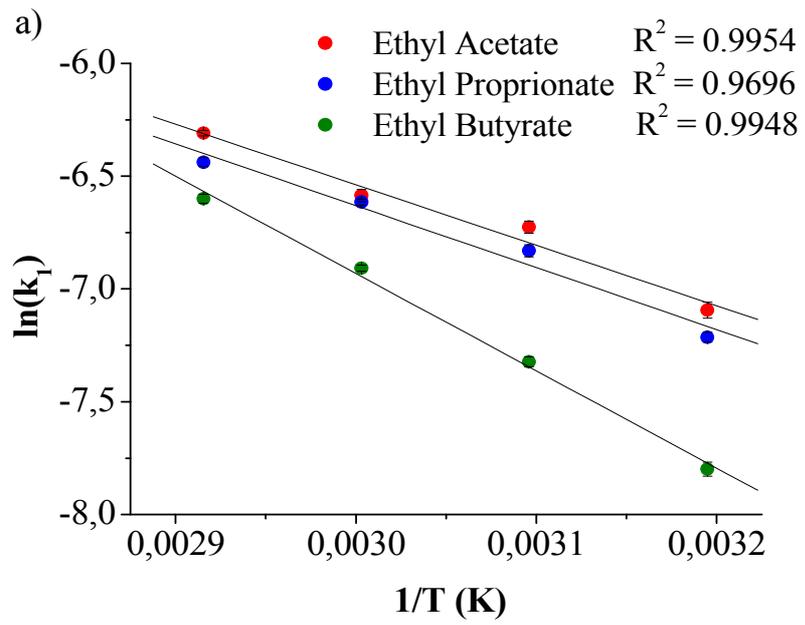

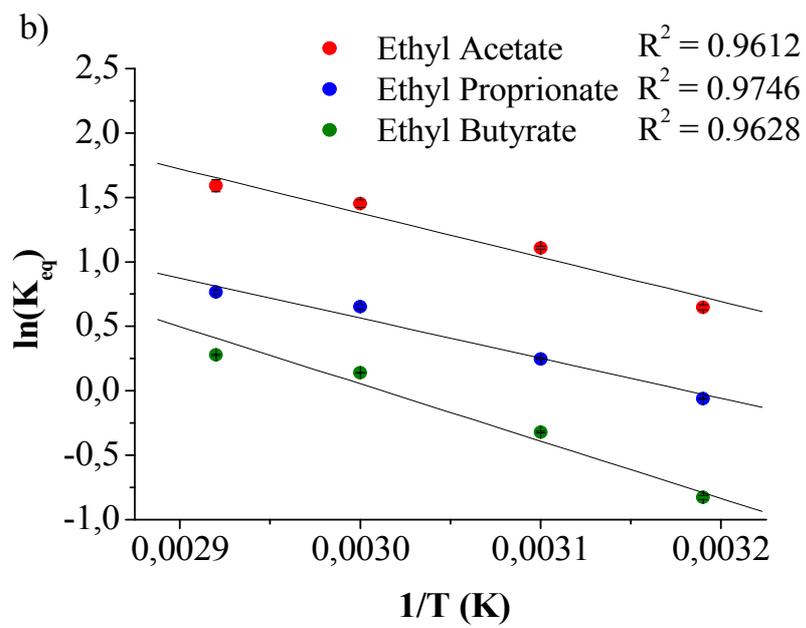

**Fig. 5**

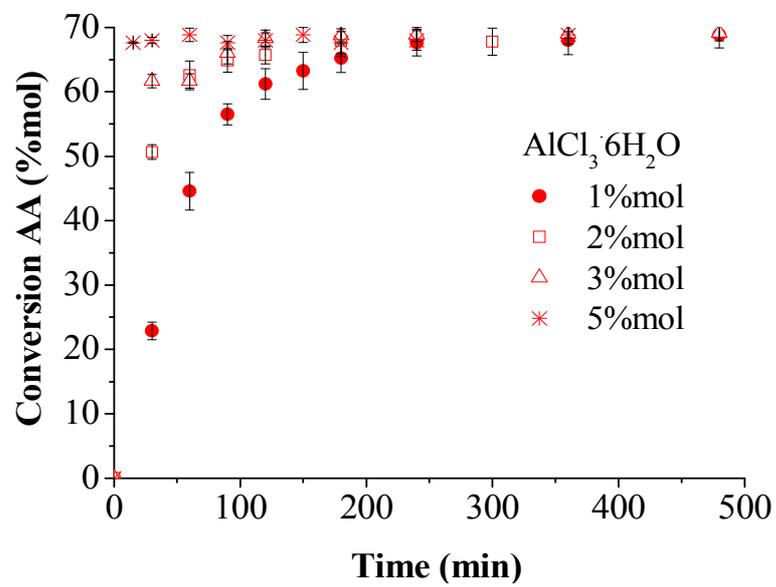

**Fig. 6**

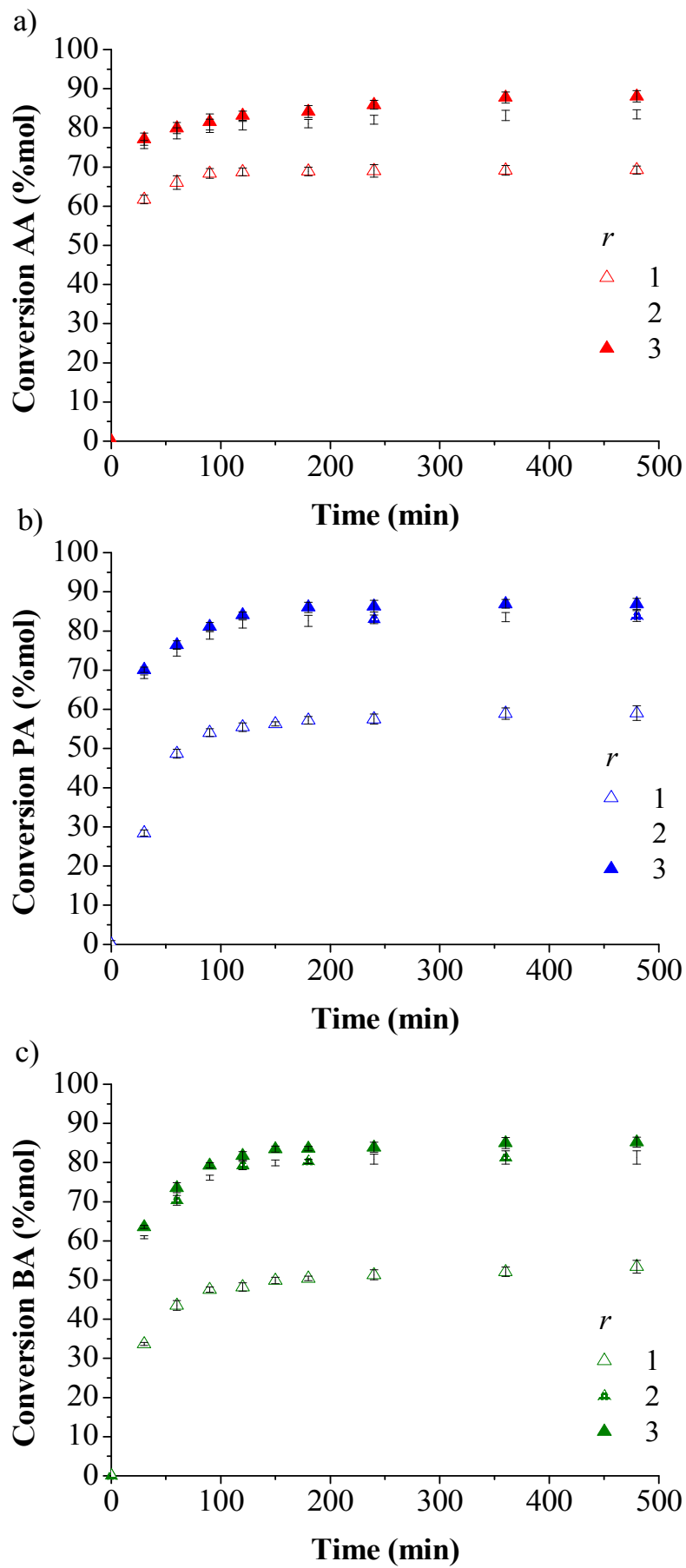

**Fig. 7**

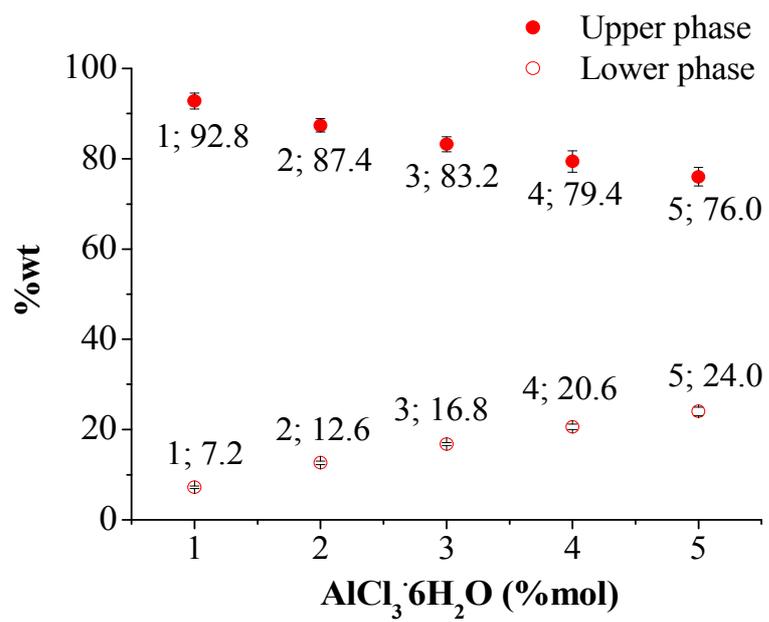

**Fig. 8**

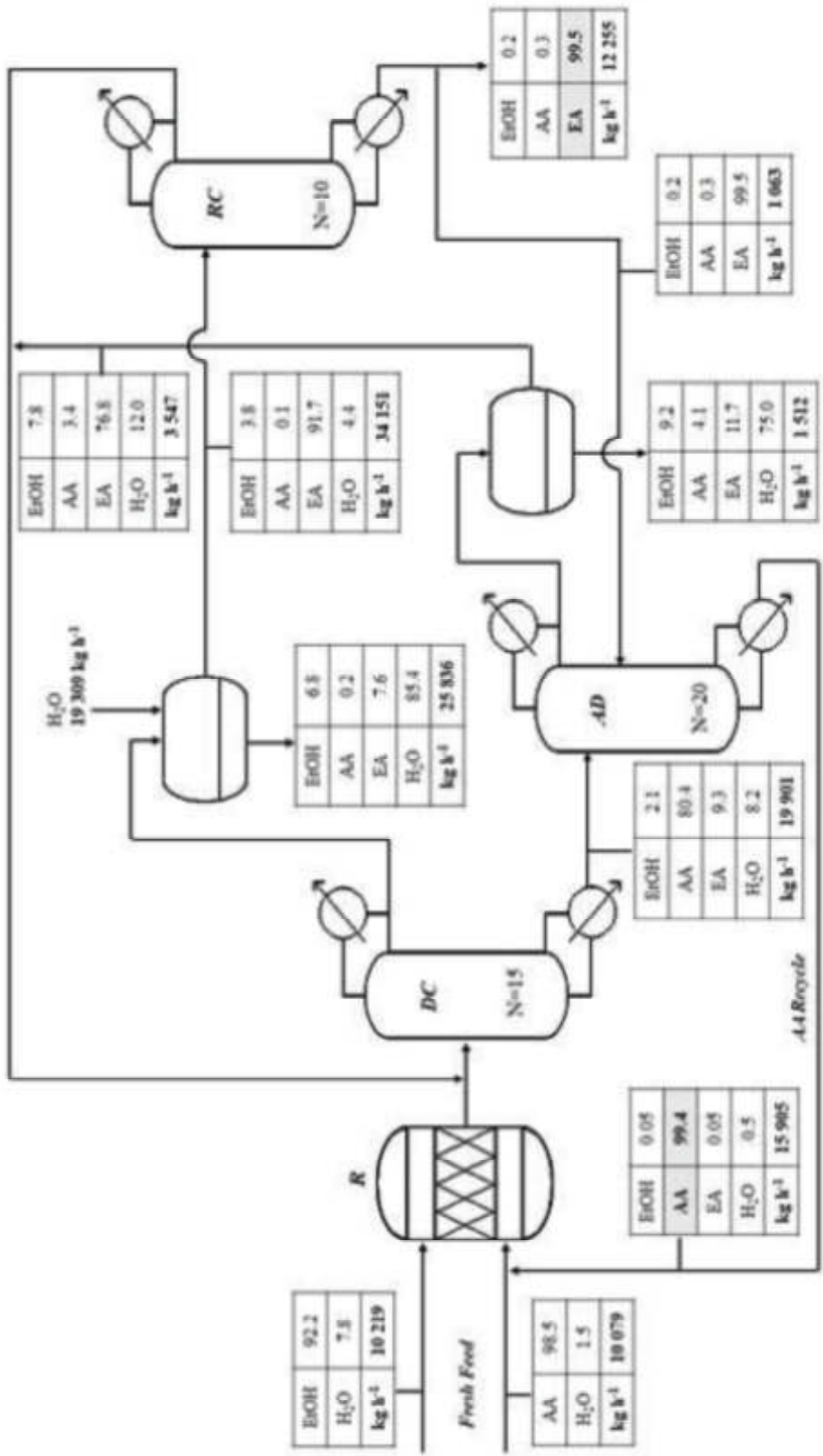

**Fig. 9**

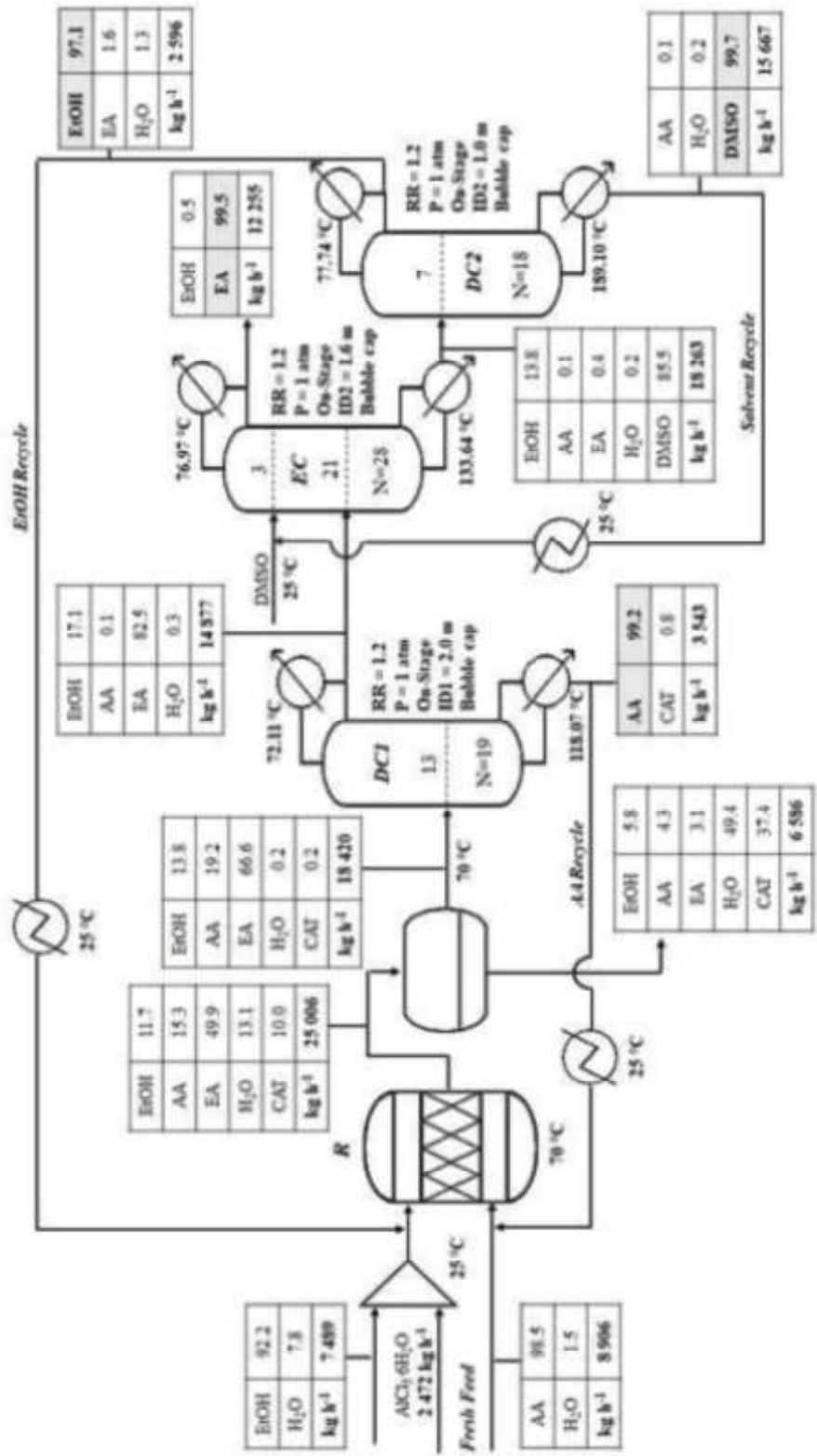